\documentclass[12pt]{iopart}
\usepackage{graphicx}
\expandafter\let\csname equation*\endcsname\undefined
\expandafter\let\csname endequation*\endcsname\undefined
\usepackage{amsmath}
\usepackage[square, sort&compress, numbers]{natbib}
\usepackage[breaklinks=true, hidelinks]{hyperref}

\usepackage{svg}
\newcommand{\AxBxC}[2]{A\textsubscript{#1}B\textsubscript{#2}C}
\newcommand{\cmVs}{cm\textsuperscript{2}V\textsuperscript{-1}s\textsuperscript{-1}}
\newcommand{\cmthr}{cm\textsuperscript{-3}}

\newcommand{\dEchy}{\Delta E_{\mathrm{c}}^{\mathrm{hy}}}

\newcommand{\AlxOy}[2]{Al\textsubscript{#1}O\textsubscript{#2}}
\newcommand{\InxAlxAs}[2]{In\textsubscript{{#1}}Al\textsubscript{{#2}}As} 
\newcommand{\InxGaxAs}[2]{In\textsubscript{{#1}}Ga\textsubscript{{#2}}As}

\newcommand{\mydiv}{\nabla\cdot}
\newcommand{\grad}{\nabla}

\newcommand{\Hlin}{H_{\mathrm{lin}}}
\newcommand{\Hnl}{H_{\mathrm{nl}}}

\newcommand{\Phiext}{\Phi_{\mathrm{ext}}}
\begin{document}

\title[Tunable Capacitor For Superconducting Qubits]{Tunable Capacitor For Superconducting Qubits Using an InAs/InGaAs Heterostructure}
\author{Nicholas Materise$^1$, Matthieu C. Dartiailh$^2$\footnote{Present address: Institut Ne\'{e}l, CNRS, France}, William M. Strickland$^2$, Javad Shabani$^2$ and Eliot Kapit$^1$}
\address{$^1$Department of Physics, Colorado School of Mines, 1500 Illinois St., Golden, CO 80401 USA}
\address{$^2$Center for Quantum Information Physics, Department of Physics, New York University, NY  10003, USA}%

\date{\today}

\begin{abstract}
Adoption of fast, parametric coupling elements has improved the performance of superconducting qubits, enabling recent demonstrations of quantum advantage in randomized sampling problems. The development of low loss, high contrast couplers is critical for scaling up these systems. We present a blueprint for a gate-tunable coupler realized with a two-dimensional electron gas in an InAs/InGaAs heterostructure. Rigorous numerical simulations of the semiconductor and high frequency electromagnetic behavior of the coupler and microwave circuitry yield an on/off ratio of more than one order of magnitude. We give an estimate of the dielectric-limited loss from the inclusion of the coupler in a two qubit system, with coupler coherences ranging from a few to tens of microseconds.
\end{abstract}
\maketitle

\section{Introduction}
Tunable couplers for superconducting qubits, previously thought of as long-term investments in future quantum computers and building blocks towards demonstrating high fidelity two qubit gates~\cite{Lu2017,Huang2018}, are now center-pieces of large scale superconducting qubit-based quantum computers. The early quantum advantage demonstration~\cite{Arute2019} owes its success, in part, to the two-qubit gate fidelities across the chip facilitated by fast, tunable couplers. Often tunable couplers are realized as mutual inductances or effective capacitances between nearest-neighbor qubits and tuned by flux-biased superconducting quantum interference devices (SQUIDs), naturally integrating with both fixed and flux-tunable superconducting qubit fabrication capabilities~\cite{Barends2019}.

Advancements in the growth of superconductor-semiconductor (super-semi) structures for use in gate-tunable Josephson junctions have led to proposals~\cite{Qi2018, Marcus2019} and experimental demonstrations of voltage-controlled coupling schemes, superconducting quantum storage units~\cite{Sardashti2020}, and readout resonator buses~\cite{Casparis2019}. Unlike their conventional transmon qubit~\cite{Koch2007} counterparts, whose energies are either fixed by their shunt capacitors or tuned with magnetic fluxes threading SQUID loops~\cite{Oliver2013}, these hybrid quantum systems consist of epitaxial III-V semiconductor layers whose properties are tunable with precise composition control and applied electric fields.

Challenges in optimizing materials and fabrication processes remain to realize high coherence gatemon~\cite{Larsen2015} qubits and other voltage-tunable super-semi devices. These gatemon qubits differ from their flux-tunable and fixed frequency transmon counterparts in that their Josephson junctions are formed by superconductor-semiconductor-superconductor junctions and their Josephson energies $E\textsubscript{J}$ are tunable by an external electric potential. Although achieving coherences of two dimensional electron gas (2DEG)-based gatemon qubits at parity with conventional transmon-like qubits remains an open area of research, similar systems acting as low participation couplers still offer fast, high contrast control with a tolerable reduction in system coherence. Recent experimental demonstrations of tunable resonators using the same materials stack, achieved an on/off coupling ratio between resonators of one order of magnitude, a promising first step towards realizing fast, voltage-tunable couplers~\cite{Strickland2023}.

We propose a voltage-controlled capacitive coupling element between neighboring superconducting qubits using a III-V semiconductor 2DEG in an InAs/InGaAs heterostructure. The capacitance of the coupler tunes as a function of a gate voltage or series of gate voltages applied to the 2DEG, repelling electrons away from the region underneath the gates. By ``parting the sea of electrons'' in the quantum well, the coupler straddles two limits -- fully conducting and fully depleted or insulating. In the intermediate region, the area of the depleted charges acts as an effective dielectric of some width $d$, and the capacitance of the coupler decreases with increasing width, as one might expect a parallel plate capacitor to behave as the separation between the plates increases. From this simple operational principle and reduction in sensitivity to bias line fluctuations, we expect such a coupler to be a drop-in replacement for SQUID-based inductive couplers~\cite{Mayer2019}.

Additional capacitors between the coupler and the qubits may minimize unwanted electric field coupling to other qubits. This is an improvement over SQUID-based couplers, where stray magnetic fields can lead to classical cross-talk between qubits~\cite{Bialczak2011}. We suspect that the 2DEG coupler may introduce more charge noise than the inductive couplers through the voltage control lines, yet transmon qubits, our initial targets for qubit-coupler integration testing, are exponentially insensitive to this charge noise.

\begin{figure}[!ht]
    \centering
    \includegraphics[width=0.35\textwidth]{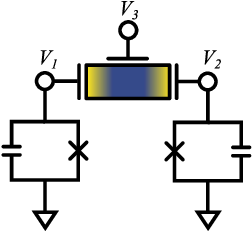}
    \caption{Schematic of two transmon qubits and the 2DEG coupler. Blue regions (color online) correspond to low electron concentration or effective dielectrics and red regions correspond to high electron concentration or effective conductors. We use the labeling of the voltage nodes $V_i$ throughout the text, where nodes 1 and 2 correspond to source and drain terminals, and node 3 refers to the gate terminal.}
    \label{fig:capacitor_qubit_schematic}
\end{figure}

The structure of the paper is as follows. We start by presenting a conceptual design of the coupler in Section~\ref{sec:conceptual_design}. In Section~\ref{sec:coupler_modeling}, we formulate rigorous numerical models of the 2DEG coupler, starting with COMSOL semiconductor electron density calculations, followed by additional electrostatic and frequency-domain COMSOL simulations of the capacitance and admittance matrices, respectively. That section concludes with a summary of the dielectric and other loss mechanisms present in the III-V semiconductor and dielectric materials in the coupler. Section~\ref{sec:int_cqed} details our ANSYS high frequency simulation software (HFSS) simulations of a prototypical two transmon qubit circuit coupled by a lumped element capacitor representing the 2DEG coupler. We apply energy participation ratio techniques~\cite{Minev2020} to extract the Hamiltonian matrix elements in the dispersive regime, and extend these calculations to compute the charge-charge interaction matrix elements between the two transmon qubits. These analyses give similar results when considering a single lumped element variable capacitor representing the coupler compared to a full parasitic capacitance model of the coupler from our electrostatic COMSOL simulations.

\section{Methods and Modeling}
%
\subsection{Conceptual Design}\label{sec:conceptual_design}
Inspired by textbook parallel plate capacitors whose capacitance varies inversely with plate separation, our coupler design relies on similar carrier dynamics as field effect transistors to electronically modify the effective parallel plate capacitor geometry seen by neighboring qubits. We consider a proximitized semiconductor~\cite{Kleinsasser1989} sandwiched between two transmon-like qubits with large capacitor plates patterned on top and a metal-oxide gate separating the two plates. Applying a negative gate voltage decreases the carrier concentration directly below the gate, modifying the capacitor geometry by increasing the effective separation of the parallel plates. The high electron mobility of the carriers in the 2DEG, exceeding 14 000 cm\textsuperscript{2}V\textsuperscript{-1}s\textsuperscript{-1} at 20 mK,~\cite{Wickramasinghe2018} allows for fast gating, enabling parametric interactions with rapidly oscillating gate voltages.

This concept generalizes to multiple gates, where each region of low electron concentration corresponds to an effective dielectric and each region with high electron concentration acts a conductor. The effective capacitance seen by the two qubits is the series combination of the individual capacitances defined by alternating effective dielectrics and conductors. Similar gating schemes have been proposed for nonreciprocal devices~\cite{Leroux2022}, tunable quantum buses~\cite{Casparis2019}, and controlled-Z gates~\cite{Qi2018}.

Apart from the aforementioned experimental demonstrations of these devices, few modeling efforts, if any, have explored the practical considerations of realizing such couplers. The following numerical simulations aim to address those concerns by estimating the capacitive tuning range in the presence and absence of parasitic capacitances, calculating relevant interaction matrix elements, and providing an upper bound on the losses inherited by the system from the dielectric materials of the coupler.

\subsection{Classical Modeling}\label{sec:coupler_modeling}
\subsubsection{Semiconductor 2DEG Calculations}
To estimate the capacitance of the 2DEG coupler, we compute the electron concentrations in the active region of the device (InGaAs/InAs/InGaAs layers) using the COMSOL Multiphysics Semiconductor Module~\cite{SemiconductorModuleUsersGuide}. Equilibrium solutions to the drift-diffusion equations with Fermi-Dirac statistics serve to identify regions of high depletion under the gate(s) when applying negative voltages on the order of a few volts, overcoming the work function of the aluminum gate contact.

\begin{figure}[!ht]
    \centering
    \includegraphics[width=0.5\textwidth]{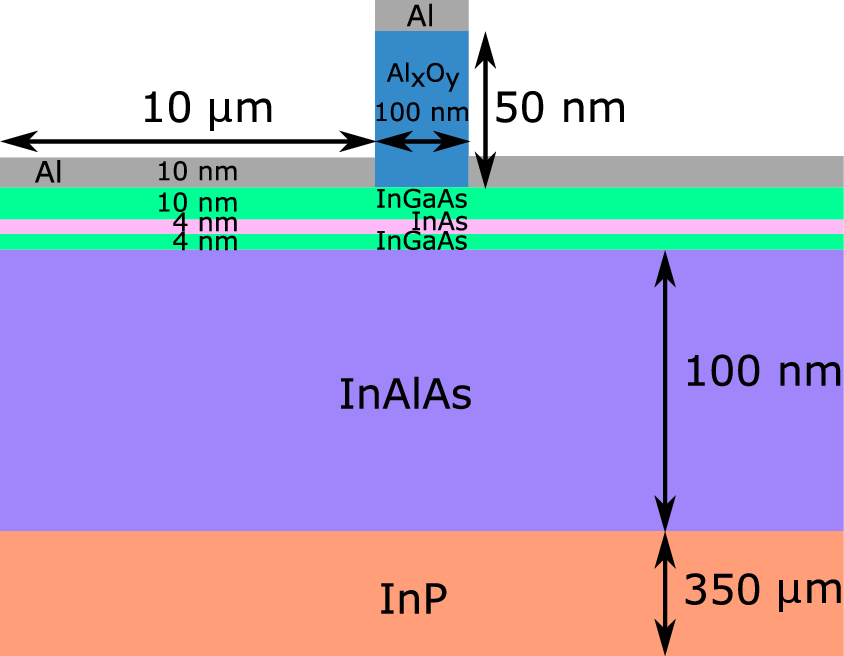}
    \caption{Schematic of the 2DEG coupler as modeled in COMSOL Multiphysics based on~\cite{Mayer2019}. An aluminum contact deposited on~\AlxOy{x}{y}~defines the gate terminal. We abbreviate the fixed composition ternary III-V alloys In\textsubscript{0.81}Ga\textsubscript{0.19}As and In\textsubscript{0.81}Al\textsubscript{0.19}As, as InGaAs and InAlAs, respectively. Not shown or modeled is the superlattice graded buffer layer between InP and InAlAs~\cite{Shabani2016, Wickramasinghe2018, Strickland2020}.}
    \label{fig:device_stack}
\end{figure}

We use a layer structure typical of gatemon qubits as in figure~\ref{fig:device_stack} and refer to this structure as the ``device stack''~\cite{Shabani2016, Wickramasinghe2018, Strickland2020, Strickland2022}. To model the device stack in COMSOL, we specified the following electronic properties of the semiconductor materials and the dielectric constant of the gate oxide: electron and hole effective conduction band masses $m^{\ast}_{\mathrm{n(p),c}}$, low-field mobilities $\mu^{\mathrm{lf}}_{\mathrm{n(p)}}$, band gap energies $E\textsubscript{g}$, conduction band offsets $\Delta E\textsubscript{c}$ between neighboring semiconductors, dielectric constants $\varepsilon\textsubscript{r}$, and effective densities of states for the conduction and valence bands $N\textsubscript{c(v)}$. Taking the electron affinity $\chi$ for InAs as given by the COMSOL material library,  we calculated the remaining affinities using Anderson's affinity rule and the conduction band offsets of each material~\cite{Streetman2015Ch5}. As in reference~\cite{Strickland2022}, we included a silicon delta-doping 6 nm below the interface between the InAlAs and lower InGasAs layers. COMSOL approximates such a doping profile with the Geometry Doping profile, which we select a Gaussian profile with a width of 0.1 nm. Table~\ref{tab:mat_table} gives a summary of the material parameters used in these semiconductor simulations; see \ref{app:iii_v_params} for detailed calculations of the energy gaps, effective masses, and conduction band offsets for \InxGaxAs{x}{1-x} and \InxAlxAs{x}{1-x} as functions of the composition parameter (x).
\begin{table}[!ht]
        \centering
        \begin{tabular}{p{0.25\linewidth}p{0.18\linewidth}p{0.18\linewidth}p{0.18\linewidth}p{0.18\linewidth}}
        \hline
        \                &  InAs  & InGaAs & InAlAs & InP\\
        \hline\
        $E\textsubscript{g}$ [eV]                      &  0.354           & 0.473            & 0.752  & 1.344 \\
        $\Delta E\textsubscript{c}$ [eV]               &  -               & 0.200            & 0.201  & 0.12\\
        $\varepsilon_{\mathrm{r}}$                 & 15.15            & 14.03            & 13.13  & 12.9\\
        $N\textsubscript{c}$ [\cmthr]             & 6.6E18 & 1.4E17  & 2.1E17 & \\
        $N\textsubscript{v}$ [\cmthr]             & 8.73E16 & 6.4E18 & 7.8E18 & \\
        $\chi$ [eV]                     & 4.9              & 4.7              & 4.5 & 4.38 \\
        $\mu^{\mathrm{lf}}_{\mathrm{n(p)}}$ [\cmVs]& 14.4E3 (500)   & 14.4E3 (450)     & 14.E3 (384) & 5.4E3 (200)\\
        $m^{\ast}_{\mathrm{n(p),c}}$ [$m_0$]     & 0.023 (1.00)     & 0.03 (0.25)      & 0.04 (0.31) & 0.08 (0.60)\\
        \hline\hline
        \end{tabular}
        \caption{Materials parameters used in the COMSOL Semiconductor Module calculations. InGaAs and InAlAs abbreviate In\textsubscript{0.81}Ga\textsubscript{0.19}As and In\textsubscript{0.81}Al\textsubscript{0.19}As. $m_0$ corresponds to the rest mass of an electron (0.511 MeV $c^{-2}$). Out-of-plane effective electron and hole masses of InAs are set to $m_0$ in the model to simulate 2DEG confinement in the xy-plane. Electron mobilities for InAs, InGaAs, and InAlAs are all set to the same value as extracted from measurements of a similar device at millikelvin temperatures~\cite{Wickramasinghe2018}. Values not in parenthesis (in parenthesis) correspond to electron (hole) properties.}
        \label{tab:mat_table}
    \end{table}

We specify the geometry in figure~\ref{fig:device_stack} using the native COMSOL CAD editor to define domains (surfaces or planes) and boundaries (lines or edges), solving for the electron density in the domains and on the boundaries. Electronic properties assigned to each domain follow from table~\ref{tab:mat_table}. We model the terminals (source -- 1, drain -- 2, gate -- 3, as in figure~\ref{fig:capacitor_qubit_schematic}) as Terminal boundary conditions with voltages $V_1$, $V_2$, $V_3$ and contact work functions $\Phi\textsubscript{c,1}$, $\Phi\textsubscript{c,2}$, $\Phi\textsubscript{c,3}$ = 4 V~\cite{SemiconductorModuleUsersGuide}.

We selected the density gradient discretization scheme~\cite{Ancona2011} in COMSOL to approximate the quantum confinement effects in the 2DEG more efficiently than a self-consistent Schr\"{o}dinger-Poisson equation calculation. The density gradients modify the equilibrium electron ($n$) and hole ($p$) concentrations by~\cite{SemiconductorModuleUsersGuide}
    \begin{eqnarray}
        n = N\textsubscript{c} F_{1/2}\left(\frac{E\textsubscript{fn} - E\textsubscript{c} + q V^{\mathrm{DG}}_{\mathrm{n}}}{k\textsubscript{B}T}\right) \label{eq:n_dg} \\
        p = N\textsubscript{v} F_{1/2}\left(\frac{E\textsubscript{v} - E\textsubscript{fp} + q V_{\mathrm{p}}^{\mathrm{DG}}}{k\textsubscript{B}T}\right) \label{eq:p_dg} \\
        N_{\mathrm{c(v)}} = \left(\frac{2 m_{\mathrm{n(p)}}^{\ast} \pi k_{\mathrm{B}} T}{h^2}\right)^{3/2}, \label{eq:N_cv}
    \end{eqnarray}
where $E_{\mathrm{c(v)}}$ is a given material's conduction (valence) band edge, $E_{\mathrm{fn(p)}}$ are the electron (hole) quasi-Fermi level energies, $F_{1/2}(\eta)$ is the Fermi-Dirac integral~\cite{Kim2019}, $k_{\mathrm{B}}$ is Boltzmann's constant, $T$ is the temperature of the device (approximate temperature of the mixing chamber stage of typical dilution refrigerators $\sim$10 mK), and $q$ is the charge of an electron or hole. The quantum potentials $V_{\mathrm{n(p)}}^{\mathrm{DG}}$ are defined in terms of the density gradients by~\cite{SemiconductorModuleUsersGuide}
    \begin{eqnarray}
        \mydiv \left(\mathbf{b}_{\mathrm{n}} \grad \sqrt{n}\right) &= \frac{1}{2}\sqrt{n}\ V_{\mathrm{n}}^{\mathrm{DG}} \label{eq:grad_n} \\
        \mydiv \left(\mathbf{b}_{\mathrm{p}} \grad \sqrt{p}\right) &= \frac{1}{2}\sqrt{p}\ V_{\mathrm{p}}^{\mathrm{DG}}, \label{eq:grad_p}
    \end{eqnarray}
with the density gradient tensors $\mathbf{b}_{\mathrm{n(p)}}$ for electrons (holes) expressed in terms of the effective mass tensors $\mathbf{m}_{\mathrm{n(p)}}^{\ast}$
    \begin{eqnarray}
        \mathbf{b}_{\mathrm{n}} = 
        \frac{\hbar^2}{12q}\left[\mathbf{m}^{\ast}_{\mathrm{n}} \right]^{-1} \label{eq:b_n} \\
        \mathbf{b}_{\mathrm{p}} = 
        \frac{\hbar^2}{12q}\left[\mathbf{m}^{\ast}_{\mathrm{p}}\right]^{-1}. \label{eq:b_p}
    \end{eqnarray}
Note the distinction between the scalar effective masses $m_{\mathrm{n(p)}}^{\ast}$ and, the effective mass tensors $\mathbf{m}^{\ast}_{\mathrm{n(p)}}$. Anisotropy in the effective mass tensors emulates the quantum confinement effects in the 2DEG, constraining electron movement to one plane.

For the remaining materials, \AlxOy{2}{3} and air, we used the Electric Charge Conservation interface, including the following constitutive relations for each dielectric in terms of its electric permittivity tensor $\pmb{\varepsilon}$~\cite{SemiconductorModuleUsersGuide}
    \begin{eqnarray}
       \mathbf{D} = \varepsilon_0\pmb{\varepsilon}:\mathbf{E}, \label{eq:D_epsE} 
    \end{eqnarray}
where $\mathbf{D}$ is the electric displacement field, $\varepsilon_0$ is the permittivity of free space, $\mathbf{E}$ is the electric field, and $\pmb{\varepsilon}:\mathbf{E}$ is a tensor contraction (matrix vector product) between $\pmb{\varepsilon}$ and $\mathbf{E}$. Modeling these regions as pure dielectrics reduces the size of the system of equations relative to a drift-diffusion calculation applied to the materials that behave as perfect insulators, air and oxide layers. We excluded the superlattice graded buffer between the InAlAs and InP, as we expect the electric fields and carrier concentrations to be negligible in those regions and the additional computational cost (number of degrees of freedom solved for in the COMSOL model) would not improve the accuracy of our estimates of the capacitances and conductances of the coupler that will be largely determined by the charge dynamics in and near the active region (InGaAs/InAs/InGaAs).
    \begin{figure}[!ht]
        \centering
        \includegraphics[width=\textwidth]{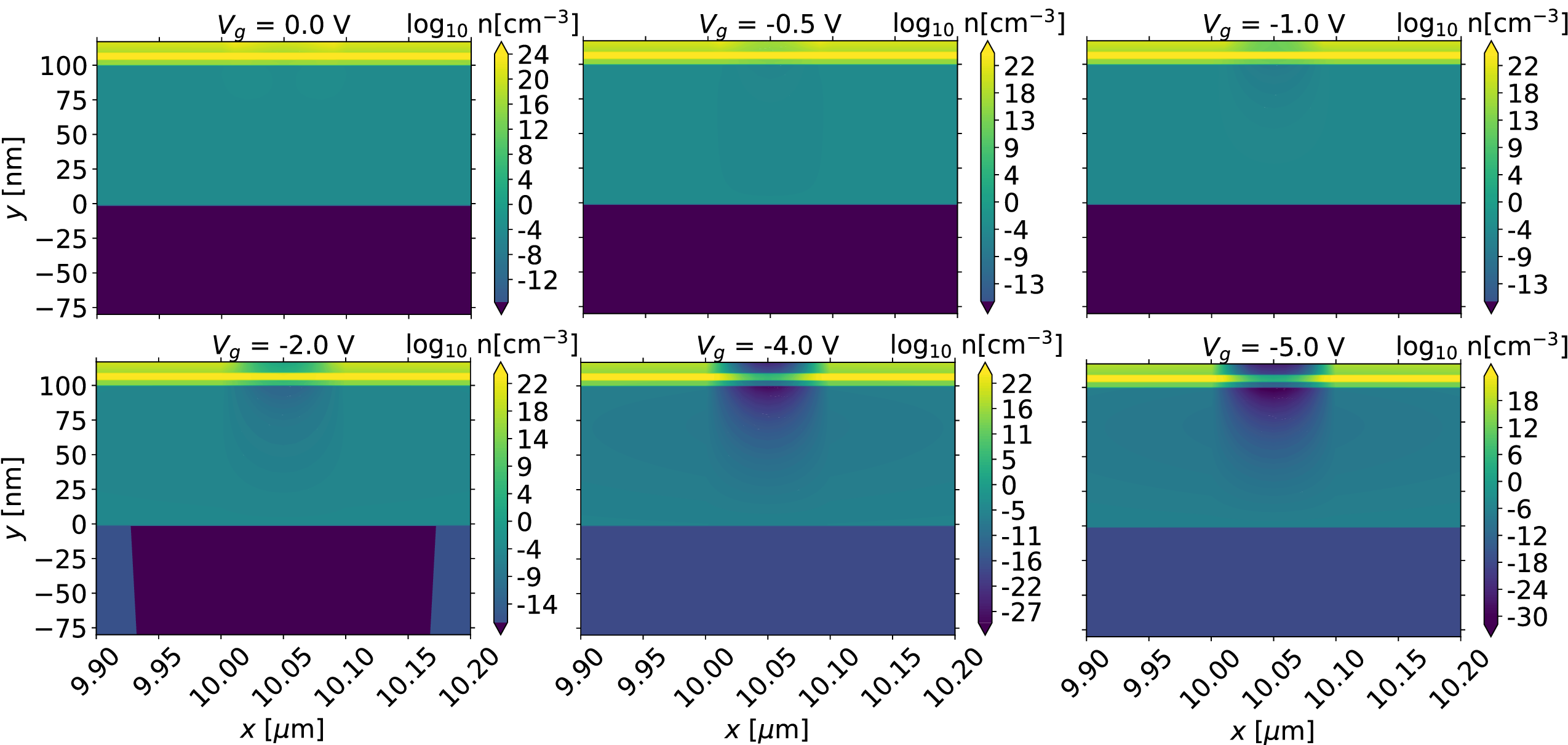}
        \caption{Electron concentrations [cm\textsuperscript{-3}] on a base-10 logarithmic scale with source-drain bias $V_{\mathrm{sd}} = 0$ V for the fully conducting $V_{\mathrm{g}} = 0$ V, intermediate $V_{\mathrm{g}}=-0.5,\,-1,\,-2,\,-4$ V,  and fully depleted $V_{\mathrm{g}}=-5$ V operating points. The horizontal axis is a 300 nm span centered on the gate electrode and the vertical axis starts at the contact--2DEG interface at $\sim$117 nm, the 2DEG--InAlAs interface is 18 nm below that, and the InAlAs--InP interface is located at 0 nm. We do not solve for $n$ in the regions where we applied the Electric Charge Conservation equations, i.e. in the \AlxOy{x}{y} regions not shown, yet the electric fields respect the boundary conditions set by those regions.}
        \label{fig:n_conducting_depleted}
    \end{figure}
%
%
\subsubsection{Electric Currents Admittance Matrix Calculations}
To extract the conductance matrix and verify the capacitance matrix of the device under high frequency excitation agrees with the electrostatic result, we use the Harmonic Perturbation option in the COMSOL Semiconductor module to compute the admittance matrix $\mathbf{Y}$ defined in terms of the $N$ terminal voltages $V_k$ and currents $I_k$~\cite{ACDCModuleUsersGuide}
    \begin{eqnarray}
        \begin{pmatrix}
        I_1 \\
        I_2 \\
        \vdots \\
        I_N
        \end{pmatrix}
        =
        \begin{pmatrix}
        Y_{11} & Y_{12} & \hdots & Y_{1N} \\
        Y_{21} & Y_{22} & \hdots & Y_{2N} \\
        \vdots & \vdots & \      & \vdots \\
        Y_{N1} & Y_{N2} & \hdots & Y_{NN}
        \end{pmatrix}
        \begin{pmatrix}
        V_1 \\
        V_2 \\
        \vdots \\
        V_N
        \end{pmatrix}
        \label{eq:def_ymatrix}
    \end{eqnarray}
In the frequency domain, the voltages and currents become phasors of the form $\widetilde{V}_ke^{i\omega t}$ and $\widetilde{I}_ke^{i\omega t}$, with the admittance matrix given by
    \begin{eqnarray}
        \mathbf{Y} = \mathbf{G} + i\omega \mathbf{C}, \label{eq:ymatrix_freq_domain}
    \end{eqnarray}
where $\mathbf{G}$ and $\mathbf{C}$ are the conductance and capacitance matrices, $i=\sqrt{-1}$, and $\omega$ is the angular frequency~\cite{ACDCModuleUsersGuide}. Both matrices are nearly symmetric for our nonlinear, three-terminal device in figure~\ref{fig:capacitor_qubit_schematic}.

The Harmonic Perturbation option applies a small AC signal with angular frequency $\omega$ to each terminal after a DC operating point has been calculated by the semiconductor solver with some voltage applied to the gate, source, and drain contacts. At each DC operating point (linearization point), COMSOL computes the currents and voltages by differentiating the perturbed solution. To compute the admittance matrix above, we compute the ratio of the current and voltage at each terminal, i.e.~\cite{SemiconductorModuleUsersGuide}
    \begin{eqnarray}
        Y_{ij} &= \frac{I_i}{V_j}\bigg|_{V_{k\neq j}=0}
        \label{eq:admittance_i_v}
    \end{eqnarray}
We compute these currents at each terminal, given voltage source excitations, as a function of frequency $\omega$ in the band of 4--8 GHz relevant to superconducting qubit and resonator frequencies, and extracted the conductance matrix as the real, frequency-independent part of $\mathbf{Y}$ and the capacitance matrix as the derivative of the imaginary part of $\mathbf{Y}$ from (\ref{eq:ymatrix_freq_domain})
    \begin{eqnarray}
        G_{ij} = \mathrm{Re}\{Y_{ij}(\omega)\} \label{eq:G_ij} \\
        C_{ij} = \mathrm{Im}\left\{\frac{\rmd Y_{ij}(\omega)}{\rmd\omega}\right\}, \label{eq:C_ij}
    \end{eqnarray}
where $\rmd Y_{ij}(\omega) / \rmd\omega$ is a constant in our case, as we omit the junction inductance $L_{\mathrm{J}_0}$ leading to discontinuities at resonance frequencies proportional to $(C_{ij}L_{\mathrm{J}_0})^{-1/2}$~\cite{Nigg2012}. The capacitance and conductance matrices $\mathbf{C}_{c(d)}$ and $\mathbf{G}_{c(d)}$ read
    \begin{eqnarray}
        \mathbf{C}_{c} [\mathrm{fF}] &=
        \begin{pmatrix}
            \phantom{-}13.2 & -13.0 & -0.17 \\
            -13.0&  \phantom{-}13.2 & -0.17 \\
            -0.17 & -0.17 & \phantom{-}0.34
        \end{pmatrix}
        \label{eq:C_c_acdc} \\
        \mathbf{C}_{d} [\mathrm{fF}] &=
        \begin{pmatrix}
            \phantom{-}15.5 & -0.32 & -0.16 \\
            -0.32 & \phantom{-}15.5 & -0.16 \\
            -0.06 & -0.06 & \phantom{-}0.33
        \end{pmatrix}
        \label{eq:C_d_acdc} \\
        \mathbf{G}_{c} [\mu\mathrm{S}] &=
        \begin{pmatrix}
            \phantom{-}23.3 & -23.3 & -6.38\mathrm{E-}4 \\
            -23.3 & \phantom{-}23.3 & -6.38\mathrm{E-}4 \\
            -6.38\mathrm{E-}4 & -6.38\mathrm{E-}4 & \phantom{-}1.28\mathrm{E-}3
        \end{pmatrix}
        \label{eq:G_c} \\
        \mathbf{G}_{d} [\mu\mathrm{S}] &=
        \begin{pmatrix}
             \phantom{-}38.3 & \phantom{-}0.532      & -3.09\mathrm{E-}4 \\
            \phantom{-}0.532 &  \phantom{-}38.3      & -3.09\mathrm{E-}4 \\
            7.74\mathrm{E-}3 & 7.74\mathrm{E-}3 & \phantom{-}6.49\mathrm{E-}4
        \end{pmatrix}
        \label{eq:G_d}
    \end{eqnarray}
We include the resistance matrices $\mathbf{R}_{c(d)}=\mathbf{G}^{-1}_{c(d)}$, for later use in Section~\ref{sec:int_cqed} where we perform coupled two qubit simulations with HFSS and reference the $R_{c(d), 12}$ matrix elements in the lumped element representation of the coupler.
    \begin{eqnarray}
        \mathbf{R}_{c}[\Omega\times 10^9] &=
        \begin{pmatrix}
            93.3 & 93.3 & 93.3 \\
            93.3 & 93.3 & 93.3 \\
            93.3 & 93.3 & 94.1 \\
        \end{pmatrix}
        \label{eq:R_c} \\
        \mathbf{R}_{d}\left[\Omega\right] &=
        \begin{pmatrix}
            \phantom{-}2.61\mathrm{E+}4 & -3.65\mathrm{E+}2 & 1.23\mathrm{E+}4 \\
            -3.65\mathrm{E+}2 & \phantom{-}2.61\mathrm{E+}4 & 1.23\mathrm{E+}4 \\
            -3.07\mathrm{E+}5 & -3.07\mathrm{E+}5 & 1.54\mathrm{E+}9 \\
        \end{pmatrix}
        \label{eq:R_d}
    \end{eqnarray}
Note that the capacitance, conductance, and resistance matrices away from the fully conducting operating point are not symmetric. The departure of the capacitance matrices from symmetry likely stems from numerical imprecision. The conductance discrepancies we attribute to the observed frequency dependence of the real part of $\mathbf{Y}$. In the symmetric setting, we expect $\mathrm{Re}\{\mathbf{Y}(\omega)\}$ to be constant with respect to $\omega$, but we find that it varies linearly with $\omega$. This frequency dependence we model as $\mathrm{Re}\{\mathbf{Y}(\omega)\} = \omega \mathbf{g}$ and report $\mathbf{G} = \mathbf{g} \Delta \omega$, where $\mathbf{g}$ has units of $\Omega^{-1}\, s$ and $\Delta\omega$ is the frequency step used by the Harmonic Perturbation study.

The matrix elements of interest, $C_{12}=C_{21},\, G_{12}=G_{21},\,\mathrm{and}\,R_{12}=R_{21}$ represent the effective capacitance, conductance, and resistance between the source and drain terminals. These terminals form capacitive contacts with any pair of qubits. The capacitance tuning ratio $r$, or on/off contrast of the 2DEG coupler is given by $r_{C}$ = $C_{c, 12}$ / $C_{d, 12}\approx 40$. Similarly, the ratio of conductances is $r_{G}$ = $G_{c, 12}$ / $G_{d, 12}\approx 43$. A back of the envelope calculation of the charge concentration in the 2DEG between the source and drain contacts, using expressions for the conductivity $\sigma=\mu_n\,e\,n$ and conductance $G_{12}/d_0 = \sigma$ gives
    \begin{eqnarray}
       n_{c, \mathrm{eff}} &= \frac{G_{c, 12}}{\mu_{\mathrm{n}}e\,d_0}
       = 2.0\mathrm{E+}17\,\mathrm{cm}^{-3} \label{eq:nc_approx} \\
       n_{d, \mathrm{eff}} &= \frac{G_{d, 12}}{\mu_{\mathrm{n}}e\,d_0}
       = 4.6\mathrm{E+}15\,\mathrm{cm}^{-3}. \label{eq:nd_approx}
    \end{eqnarray}
These \textit{effective} charge concentrations agree with figure~\ref{fig:n_conducting_depleted} and link the change in capacitance and conductance with a change in carrier concentration between the source and drain contacts.
%

\section{Results}
\begin{figure}[!ht]
    \centering
    \includegraphics[width=\textwidth]{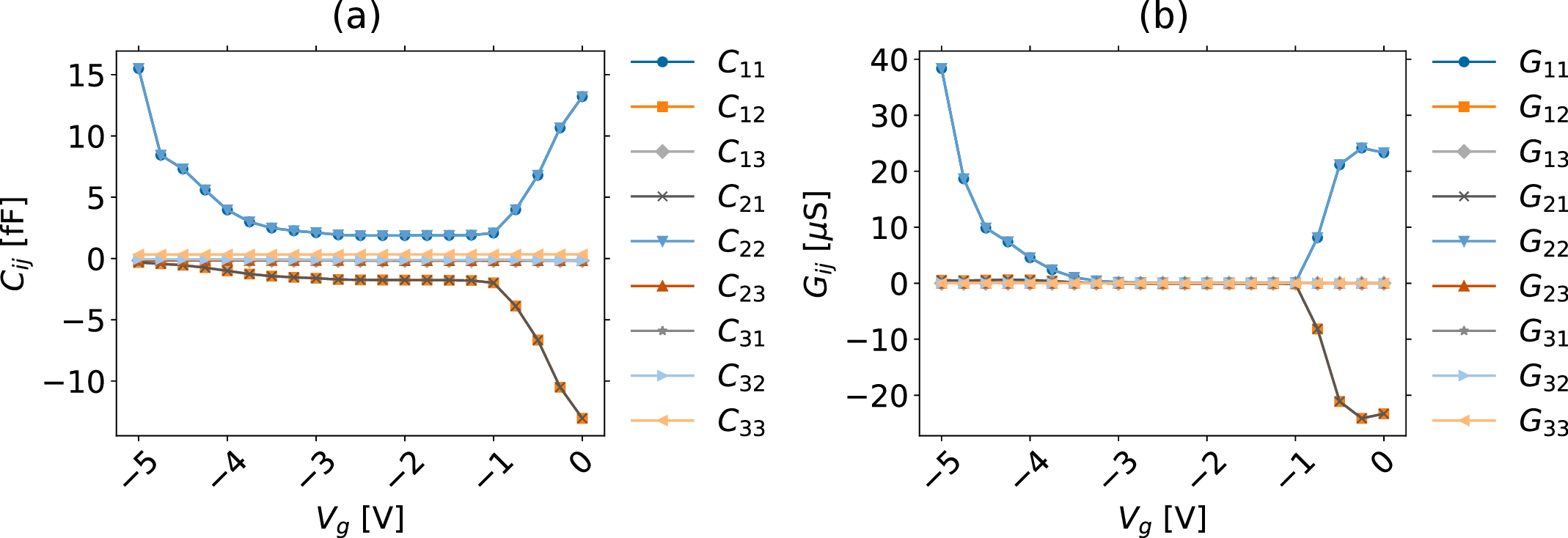}
    \caption{(a) Maxwell capacitance and (b) conductance matrices as computed with the Harmonic Perturbation study of the COMSOL Semiconductor Interface as a function of the gate voltage $V_g$.}
    \label{fig:maxwell_capacitance_matrix_vs_vg}
\end{figure}

We report the gate voltage dependence of the capacitance and conductance matrices computed in the previous section at intermediate DC operating points between the fully conducting ($V_g=0\,\mathrm{V}$) and fully depleted ($V_g=-5\,\mathrm{V}$) limits. Figure~\ref{fig:maxwell_capacitance_matrix_vs_vg} shows that the source-drain capacitance $C_{12}=C_{21}$ saturates quickly, as $V_g<-1\,\mathrm{V}$. This is a desired feature for practical tunable couplers, as lower operating voltages are preferred to reduce the active heat load from DC control signals~\cite{Krinner2019}. The conductances $G_{12}=G_{21}$ follow a similar trend, with the other matrix elements following a different $V_g$--dependence than the capacitance matrix elements.
%
\subsection{Coupler Loss Estimates}
We give bounds on the losses introduced by the 2DEG coupler from experimental measurements of the high participating gate dielectrics and InGaAs upper layer, along with the other layers in the device stack. In table~\ref{tab:pratios}, we compute the electric field participation ratios $p_j$ following the procedure developed by~\cite{Wenner2011, Wang2015}. The relaxation time $T_1$ at a given angular frequency $\omega$, as a function of the dielectric material properties and geometric factors, reads~\cite{Wang2015}
    \begin{eqnarray}
        T_1^{-1} &= \frac{\omega}{Q} = \omega\sum_{j}\frac{p_j}{Q_j} + \Gamma_0 \label{eq:T1_participation} \\
        Q_j^{-1} &= \tan\delta_j \label{eq:Qj_tan_delta_j} \\
        p_j &= W_{\mathrm{e}}^{-1}t_{\mathrm{oxide}}\varepsilon_{1,j}
        \int_{S_j}\left|\mathbf{E}\right|^2 dS \label{eq:pj} \\
        W_{\mathrm{e}} &= \int_{V} |\mathbf{E}|^2 dV \label{eq:We}
    \end{eqnarray}
where $W_{\mathrm{e}}$ is the electric field energy density stored in the volume of the entire geometry $V$, $Q_j$ are the quality factors, $\tan\delta_j$ are the loss tangents, $\varepsilon_{1,j}$ are the real parts of the dielectric function, and $t_{\mathrm{oxide}}$ is the thickness of the participating lossy surface, assumed to be 3 nm for all materials~\cite{Wang2015}. The participation ratios give the fraction of the electrical energy stored in a given surface $S_j$ relative to the total electrical energy stored in the entire volume  of the device. The last term in (\ref{eq:T1_participation}), $\Gamma_0$, includes all other loss mechanisms contributing to $T_1$ besides dielectric loss~\cite{Wang2015}. Note, these participation ratios differ from those in subsequent calculations involving \textit{energy} participation ratios referenced to a given mode rather than a particular surface.

Other sources of loss relevant to III-V semiconductor materials, but not considered in this study, include piezoelectricity~\cite{Scigliuzzo2020}, non-equilibrium quasiparticles~\cite{Nguyen2022}, cosmic ray muon flux~\cite{Wilen2021}, and, to a lesser extent, stray magnetic fields~\cite{Kringhoj2021}.
\begin{table}[ht]
        \centering
        \begin{tabular}{lllll}
            \hline\hline
            Depleted & $t_j$ [nm] & $p_{j,\mathrm{norm}}$ & $\tan\delta_j^{\ast}$ & $T_1$ [$\mu$s] \\
            \hline
            InGaAs (Top)               & 10    & 4.19E-2 & 4.1E-4   & 1.85E+0  \\
            InAs                       & 4     & 1.03E-2 & 4.1E-4   & 7.53E+0  \\
            InGaAs (Bottom)            & 4     & 7.85E-3 & 4.1E-4   & 9.89E+0  \\
            InAlAs                     & 100   & 2.92E-2 & 4.1E-4   & 2.67E+0  \\
            \AlxOy{2}{3}~\cite{McRae2020b} & 50    & 9.04E-1 & 5E-3 & 6.87E-3 \\
            InP                        & 3.5E3 & 6.97E-3 & 4.1E-4   & 8.95E+0  \\
            Total                      & -     & 1       & 7.3E-3   & 6.81E-3  \\
            \hline
            Conducting & $t_j$ [nm] & $p_{j,\mathrm{norm}}$ & $\tan\delta_j^{\ast}$ & $T_1$ [$\mu$s] \\ 
            \hline
            InGaAs (Top)                & 10    & 1.01E-8 & 4.1E-4     & 7.69E+6 \\
            InAs                        & 4     & 3.73E-9 & 4.1E-4     & 2.08E+7 \\
            InGaAs (Bottom)             & 4     & 4.03E-9 & 4.1E-4     & 1.93E+7 \\
            InAlAs                      & 100   & 1.10E-9 & 4.1E-4     & 7.06E+7 \\
            \AlxOy{2}{3}~\cite{McRae2020b}  & 50    & 9.9999E-1 & 5E-3 & 6.24E-3 \\
            InP                         & 3.5E3 & 1.41E-5 & 4.1E-4     & 4.43E+3 \\
            Total                       & -     & 1       & 7.3E-3     & 6.24E-1 \\
            \hline\hline
        \end{tabular}
        \caption{Participation ratios $p_j$, dielectric loss tangents $\tan\delta_j$, layer thicknesses $t_j$, and estimated dielectric-loss-limited $T_{1,j}$. All $T_{1,j}$ times are referenced to a qubit frequency of $\omega/2\pi$ = 5 GHz and $\tan\delta_j^{\ast}$ indicates that in the absence of reliable loss tangent data for the individual InAs, InGaAs, InAlAs, and InP layers, we used the low power loss extracted from measurements of an Al patterned CPW resonator on the full III-V stack modeled in this work and measured at 100 mK as an estimate~\cite{Phan2022}.}
        \label{tab:pratios}
\end{table}
%
\subsection{Integration with Circuit QED}\label{sec:int_cqed}
\subsubsection{Two Qubit Coupler}
\begin{figure}[!ht]
    \centering
    \includegraphics[width=0.6\textwidth]{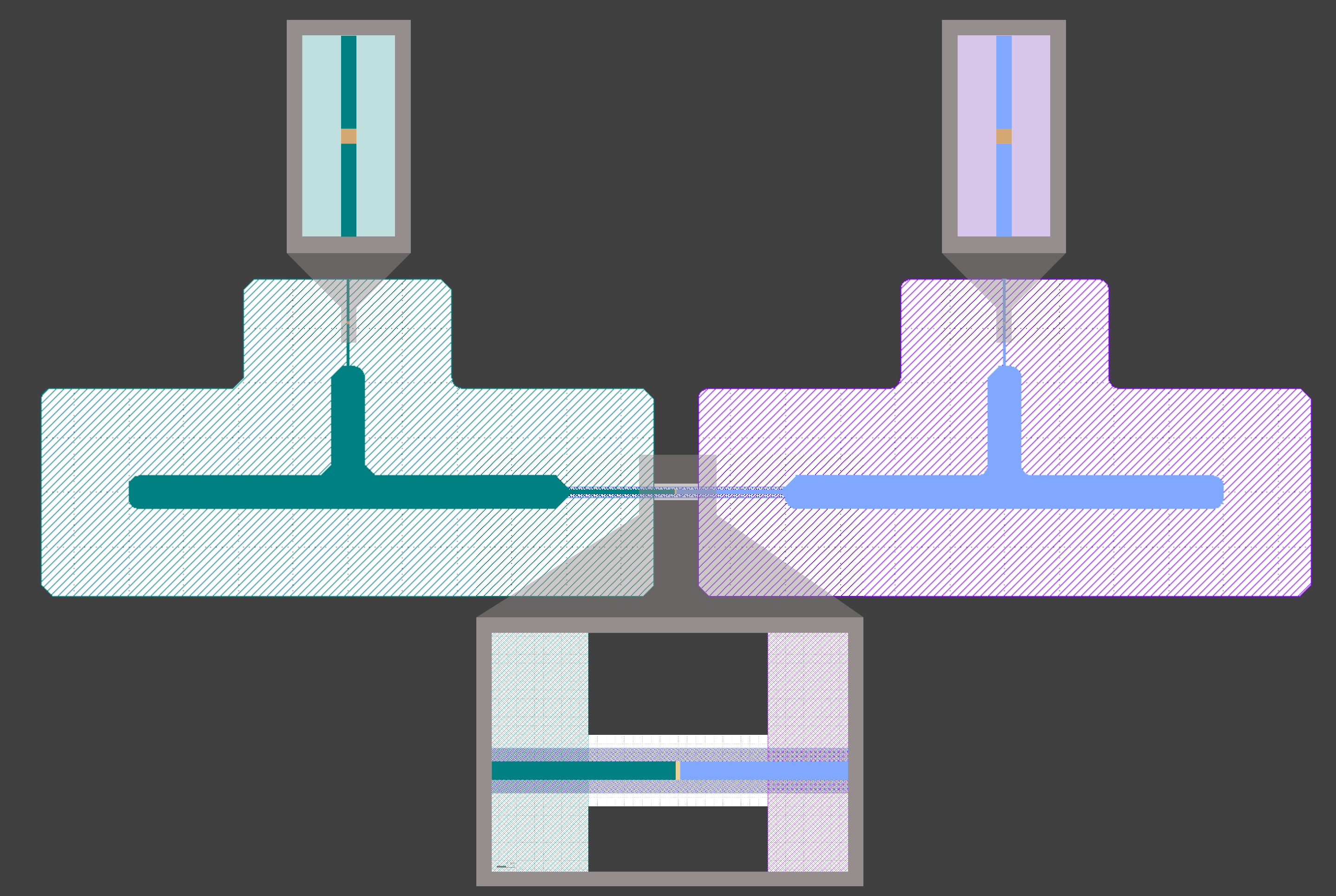}
    \caption{False color geometry of two transmon qubits with the capacitive coupler in between used in the HFSS simulations. Lumped impedances defined in the gold regions of the insets, represent the linear response of the Josephson junctions and capacitive coupling element in the HFSS model.}
    \label{fig:two_qubit_coupler}
\end{figure}

In figure~\ref{fig:two_qubit_coupler} we have a microwave circuit model of two transmon qubits coupled by a lumped impedance $Z_{\mathrm{JJ}_{\mathrm{c}}}(\omega) = 1/\left(1/R + i\omega C\right)$, where $R$ and $C$ take the values of $R_{12}$ and $C_{12}$ in either the fully conducting or fully depleted limits of the 2DEG coupler. In the conducting limit, where some current can flow across the coupler and act like a Josephson junction, one might consider adding an inductance to the coupler lumped element model. Taking $R_{12}$ to be the normal resistance of a Josephson junction and computing the junction inductance with the Ambegaokar-Baratoff formula~\cite{Ambegaokar1963}, the junction inductances would be very small, on the order of a few aH to tens of fH, resulting in high coupler mode frequencies, far outside of the frequency band of the finite element electromagnetic field solver, Ansys HFSS. For this reason and expected small modifications to qubit-qubit interactions, we omit these inductances in our model and use Ansys HFSS to compute the lowest electromagnetic eigenmodes of the device with the two transmon qubits, indexed by $j$, defined as parallel $LC$ lumped elements, $Z_{\mathrm{JJ}_{\mathrm{q}},j} = 1/\left(1/(i\omega L_{\mathrm{q}, j}) + i\omega C_{\mathrm{q},j}\right).$ In the following section, we use these eigenmode solutions to estimate the Hamiltonian matrix elements corresponding to qubit-qubit mode and qubit-coupler mode coupling strengths. We will differentiate between this modal coupling from direct capacitive coupling in the final part of this section, where we calculate the direct charge-charge interaction matrix elements.
%
\subsubsection{Energy Participation Ratios and Quantization}
To extract the coupling matrix elements between the qubits in our microwave device layout, we employ the energy participation ratio (EPR) method developed by Minev~\cite{Minev2019}. This approach goes beyond the larger family of black box quantization methods~\cite{Nigg2012, Solgun2014}, where the Hamiltonian describes a collection of Josephson junction-based qubits interacting with any number of harmonic modes separates into linear and nonlinear terms.

One can relate the modal decomposition of the classical electromagnetic response, e.g. impedance, admittance, or electromagnetic energies, with the linear parts of the Hamiltonian. Additional inputs describing the Josephson junction energy scales, $E_{\mathrm{J}}$ and $E_{\mathrm{C}}$, related to the inductive and capacitive energies of the junction, account for the nonlinear terms. The total Hamiltonian, accounting for $M$ modes, in the dispersive regime and under the rotating wave approximation, reads
    \begin{eqnarray}
        H &= \Hlin + \Hnl \label{eq:H_disp} \\
        \Hlin / \hbar &= \sum_{m=1}^M\omega_m a_m^{\dagger}a_m \label{eq:Hlin} \\
        \Hnl / \hbar &= -\sum_{m=1}^M\left(\Delta_ma_m^{\dagger}a_m  + \frac{1}{2}\alpha_m 
        a_m^{\dagger\ 2}a_m^2 \right) \nonumber \\
        &+ \frac{1}{2}\sum_{m\neq n} \chi_{mn} a_m^{\dagger} a_m a_n^{\dagger}a_n, \label{eq:Hnl}
    \end{eqnarray}
where the Lamb shifts $\Delta_m$, cross-Kerr coefficients $\chi_{mn}$, and anharmonicities $\alpha_m$ are given by~\cite{Minev2019}
    \begin{eqnarray}
        \Delta_m &= \frac{1}{2}\sum_{n=1}^{M}\chi_{mn} \label{eq:lamb_shift} \\
        \chi_{mn} &= -\sum_{j\in J}\frac{1}{2} \frac{\hbar\omega_m\omega_n}{4E_{\mathrm{J}_j}} \label{eq:chi_mn} \\
        \alpha_m &= \frac{1}{2} \chi_{mm}. \label{eq:alpha_m}
    \end{eqnarray}

The cross- and self-Kerr (anharmonicities) coefficients extracted with the pyEPR Python package~\cite{Minev2019, Minev2020} are given by the entries of the $\pmb{\chi}_{c(d)}$ matrix in the conducting (c) and depleted (d) limits of the coupler
    \begin{eqnarray}
        \frac{1}{2\pi}\pmb{\chi}_c \ [\mathrm{MHz}] &= \begin{pmatrix}
           223 & 67.1 \\
           67.1 & 223 
        \end{pmatrix}
        \label{eq:chi_c}\\
        \frac{1}{2\pi}\pmb{\chi}_d \ [\mathrm{MHz}] &= \begin{pmatrix}
           129 & 1.02 \\
           1.02 & 129
        \end{pmatrix}
        \label{eq:chi_d}
    \end{eqnarray}
The rows and columns of $\pmb{\chi}_{c(d)}$ correspond to qubits 1 and 2. Note that the diagonal entries include the $1/2$ factor in the definition of the anharmonicities as in~(\ref{eq:alpha_m}). The eigenfrequencies and quality factors are recorded in table~\ref{tab:epr_chi_w_alpha} and follow from the HFSS eigenmode solutions.
%
%
    \begin{table}[!ht]
        \centering
        \begin{tabular}{lll}
          \hline\hline
          Qubit Index & $\omega/2\pi$ [GHz] & Q \\
          \hline
          1 (d) &  6.0228 & 1.7E7  \\
          2 (d) &  8.6135 & 5.1E9  \\
          \hline
          1 (c) &  6.0228 & 4.5E8 \\
          2 (c) &  8.6135 & 1.3E9 \\
          \hline\hline
        \end{tabular}
        \caption{Eigenmode frequenices and quality factors computed with HFSS in the conducting (c) and depleted (d) limit of the coupler.}
        \label{tab:epr_chi_w_alpha}
    \end{table}
    \begin{figure}[!ht]
        \centering
        \includegraphics[width=0.5\textwidth]{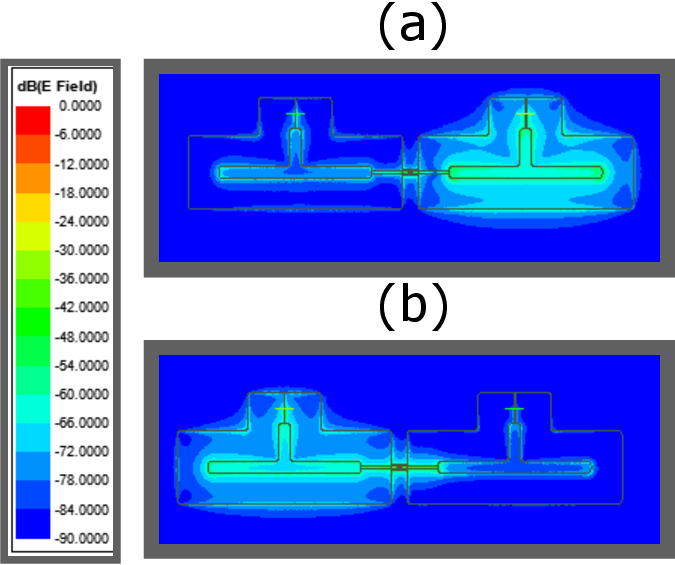}
        \caption{Electric field magnitude (dB scale to enhance color contrast) for the first two eigenmode solutions computed with HFSS. (a) 6.0228 GHz and (b) 8.6135 GHz qubits in the fully depleted limit of the coupler.}
        \label{fig:hfss_eigenmodes}
    \end{figure}
%
%
\subsubsection{Extraction of the Exchange Interaction}
To compute the charge-charge interaction strength between the transmon qubits in our HFSS model, we consider the capacitance matrix associated with a persistent current or flux qubit following the derivation by Orlando~\cite{Orlando1999}. For details on the derivation of the capacitance matrix, see \ref{app:charge_matrix}. The Hamiltonian for the coupled two transmons, written in terms of the Josephson junction phases $\varphi_j$ and node charges $q_j$, is given by
    \begin{eqnarray}
        H &= \frac{1}{2}\mathbf{Q}^T\mathbf{C}^{-1}\mathbf{Q} + U(\mathbf{\varphi}) \label{eq:def_H_charge_phase}\\
        U(\mathbf{\varphi}) &= \sum_jE_{\mathrm{J}_j}\left(1 - \cos\varphi_j\right) \label{eq:def_U_phase} \\
        \mathbf{C} &=
        \begin{pmatrix}
        C_1 + C_3 & -C_3 \\
        -C_3 & C_2 + C_3
        \end{pmatrix}. \label{eq:def_c_matrix_qq_main}
    \end{eqnarray}

In (\ref{eq:def_H_charge_phase}), the charge-charge matrix elements are one half the entries of the inverse of the capacitance matrix. We numerically inverted $\mathbf{C}$ in (\ref{eq:def_c_matrix_qq_main}) using values for $C_1$, $C_2$ obtained from (\ref{eq:alpha_to_CJ}) and $C_3=C_{12}(V_{\mathrm{g}})$ in the depleting and conducting limits. This matrix $\frac{1}{2}e^2\mathbf{C}_{c(d)}^{-1}$ is given by
    \begin{eqnarray}
       \frac{1}{2}e^2\mathbf{C}_{c}^{-1} [\mathrm{MHz}] &= \begin{pmatrix}
           179 & 21.3 \\
           21.3 & 179
       \end{pmatrix}
       \label{eq:Cinv_d} \\
       \frac{1}{2}e^2\mathbf{C}_{d}^{-1} [\mathrm{MHz}] &= \begin{pmatrix}
           200 & 0.66 \\
           0.66 & 200 
       \end{pmatrix}
       \label{eq:Cinv_c}
    \end{eqnarray}
The ratio of the off diagonal elements in (\ref{eq:Cinv_d}) and (\ref{eq:Cinv_c}) recovers an on/off interaction ratio of more than one order of magnitude, $r_{\mathrm{int}}\approx 32$.

We emphasize here that the off-diagonal charge-charge interaction matrix elements give a more accurate description of the coupling between the qubits mediated by the 2DEG coupler than the EPR calculations of the cross-Kerr coefficients.

\begin{figure}[!ht]
    \centering
    \includegraphics[width=\textwidth]%
    {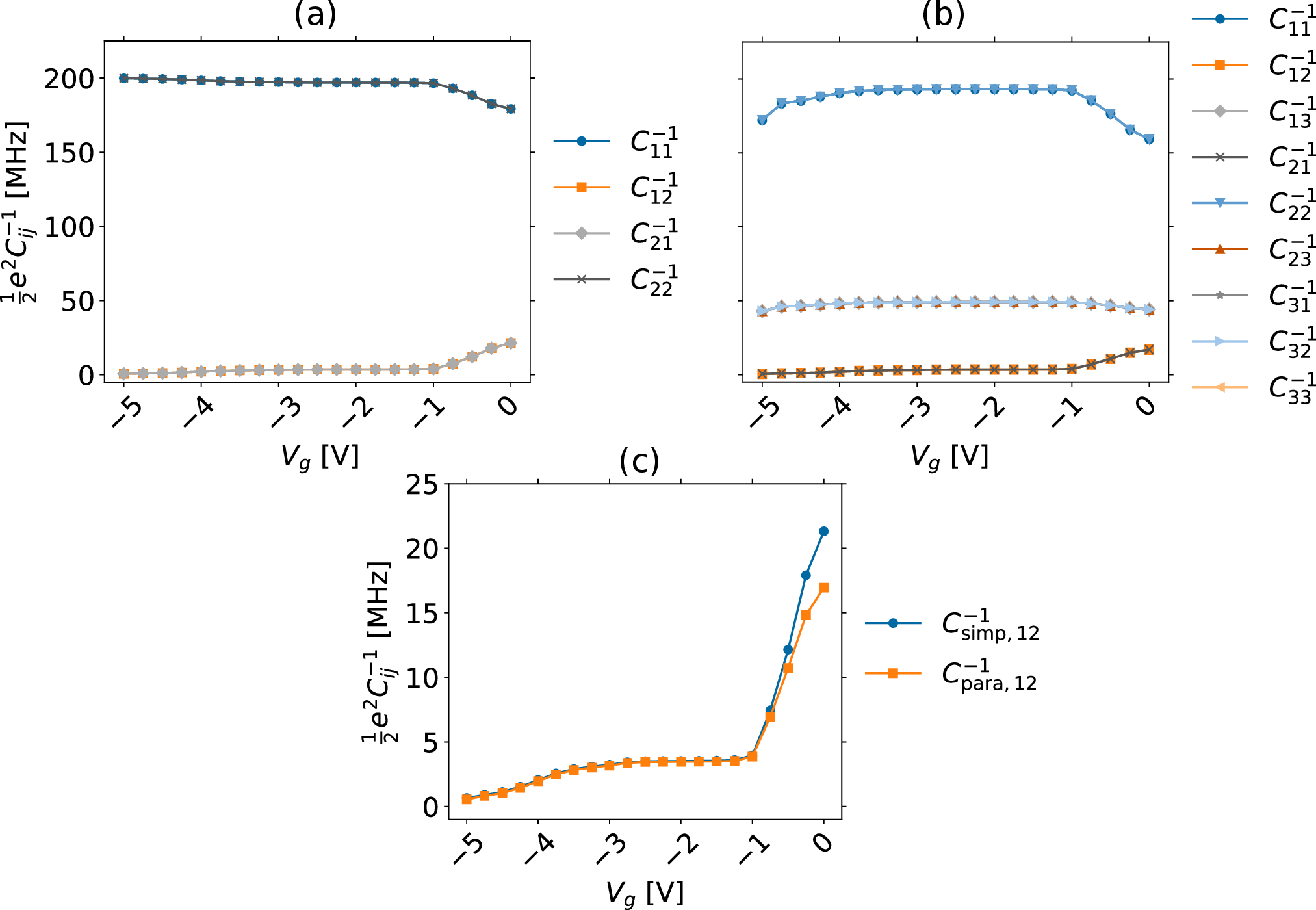}
    \caption{Coupling matrix elements in (a) the simplified two node model, (b) the three node parasitic capacitance model, (c) direct comparison of the 1-2 matrix element representing the charge-charge exchange rate between qubits 1 and 2 with the parasitic (para) and simplified (simp) capacitance matrices.}
    \label{fig:para_simp_matrix_elements}
\end{figure}

A more detailed treatment of the coupler including the parasitic capacitances from the Maxwell capacitance matrices in (\ref{eq:C_c_acdc}) and (\ref{eq:C_d_acdc}) give three-by-three coupling capacitance matrices in the Lagrangian of the form in (\ref{eq:cmatrix_parasitic}). The modified charge-charge interaction matrix elements in the full parasitic capacitance model is given by
    \begin{eqnarray}
        \frac{1}{2}e^{2}\mathbf{C}_{c,\mathrm{para}}^{-1} [\mathrm{MHz}] &=
        \begin{pmatrix}
            159 & 16.9 & 44.1 \\
           16.9 & 108  & 44.1 \\
           44.1 & 44.1 & 2.86\mathrm{E+}4
        \end{pmatrix} \label{eq:Cinv_para_c} \\
        \frac{1}{2}e^{2}\mathbf{C}_{d,\mathrm{para}}^{-1} [\mathrm{MHz}] &=
        \begin{pmatrix}
            172 & 0.552 & 43.0 \\
            0.552 & 113 & 43.0 \\
            43.0 & 43.0 & 2.99\mathrm{E+}4
        \end{pmatrix} \label{eq:Cinv_para_d}
    \end{eqnarray}
In the parasitic capacitance model, we find an on/off interaction ratio of $r_{\mathrm{int}}\approx 31$. Figure~\ref{fig:para_simp_matrix_elements} illustrates the excellent agreement between the simplified and parasitic capacitances as a function of the gate voltage. Although the simplified model does not account for the parasitic capacitances $C_{13},\,C_{31},\,C_{23},\,\mathrm{and}\,C_{32}$, it captures the behavior of the charge-charge exchange matrix elements accurately, as the parasitic contributions do not significantly change the values of $C^{-1}_{12}$.
%
\subsection{Estimation of Coupler Coherence Limit}
To estimate the total coherence limit of our coupler in the two qubit device in figure~\ref{fig:hfss_eigenmodes}, a back-of-the-envelope calculation of the energy stored in the coupler surface, in either qubit mode, gives an electric field participation ratio on the order of 10\textsuperscript{-3}, resulting in a coherence limit of a few to tens of $\mu$s, when considering the loss to be dominated by the gate dielectric and the top InGaAs layer. This is consistent with previous studies of transmon qubits whose participations are near unity in their given mode~\cite{Nigg2012,Minev2020} and gives us further confidence that a coupler of a similar geometry could support fast parametric operations with moderate coherence.

\section{Discussion}\label{sec:discussion}
From our COMSOL simulations of the 2DEG semiconductor physics, electrostatic and electric current analyses, we modeled a tunable capacitor with an order of magnitude on/off contrast. The numerical results agree with the schematic picture of modulating a parallel plate geometry by gating a high mobility 2DEG. At the level of estimating the lumped capacitance and resistance inputs to HFSS, our models incorporate 2D semiconductor behavior in greater detail than previous mixed experimental/computational reports~\cite{Wickramasinghe2018, Mayer2019}.

Two models of the coupler, with and without the parasitic capacitances extracted from COMSOL, give similar on/off interaction ratios and absolute interaction strengths. By simulating the full capacitance matrix of the multi-terminal coupler device, we motivating the choice of single gate over multi-gate coupler designs~\cite{Elfeky2021}. Both the simplified and parasitic capacitance estimates of the interaction strengths fall between hundreds of kHz to tens of MHz of coupling, on the same order of magnitude as flux-tunable couplers~\cite{Bialczak2011, Arute2019}.

Our coherence estimates further emphasize that incorporating our coupler design with existing transmon qubit designs comes at a modest reduction in system coherence. With coupler coherences limiting the system coherence to tens of $\mu$s and expected improvements in the base coherences of the coupler materials, we are optimistic that future couplers using a similar operational principle as 2DEG coupler may incur a lower system coherence penalty with the same low participation as modeled here.

This work has implications in the quantum annealing context, where both inductive and capacitive coupling may lead to nonstoquastic Hamiltonians, those that cannot be simulated by quantum Monte Carlo techniques due to the sign problem~\cite{Ozfidan2020}. By coupling conjugate degrees of freedom, charge and flux, gate-based superconducting qubit systems also stand to benefit from a richer native gate set, e.g. XX, YY, and ZZ~\cite{Consani2020, HitaPerez2021}. 

\section{Conclusion}\label{sec:conclusion}
We simulated a 2DEG-based, voltage-controlled tunable coupler compatible with superconducting qubits. With an estimated tuning ratio of one order of magnitude and similarly reduced conductance in the off state, our proposed design is an excellent candidate for capacitively coupling superconducting qubits. HFSS simulations and subsequent capacitance matrix inversion analysis suggest that the coupling matrix elements exhibit the same range of tunability, and estimates of the loss suggest that the gate oxides limit the lifetime of the coupler to nearly ten $\mu$s and the top InGaAs layer limits coupler lifetimes to several tens of $\mu$s. Low loss gate dielectrics such as tantalum oxide~\cite{Dutta2022} and hexagonal boron nitride~\cite{Wang2022}, along with improvements in the fabrication of the III-V stack may increase these coherence limits in the near term. Subsequent design iterations may look to reduce the parasitic capacitances with geometric optimization techniques to maximize contrast and minimize stray interactions~\cite{Holland2017}. A tunable capacitive element may also can serve a complementary role with tunable inductive elements to realize nonstoquastic Hamiltonians in quantum annealing systems~\cite{Ozfidan2020}. Our tuning and coherence estimates, coupled with the benefit of exponential suppression of charge noise over first order sensitivity flux noise in SQUID-based couplers, give us confidence that voltage-controlled coupling elements of the form developed here have the potential to supplant and complement their inductive counterparts in superconducting qubit systems.

\section*{Acknowledgments}
We acknowledge funding from NSF Grant PHY-1653820 and Army Research Office Grant Nos. W911NF-18-1-0125 and W911NF-18-1-0067. N. M. acknowledges funding from the Graduate Fellowships for STEM Diversity and W. M. S. acknowledges funding from the Army Research Office QuaCGR Fellowship. We would like to thank Bradley Llyod, Chien Liu, Paul Niyonkuru, Alan Phillips, John Rose, Meenakshi Singh, and David Young for many insightful discussions.

\section*{Data Availability}
The data that support the findings of this study are openly available at the following DOI: \href{https://doi.org/10.5281/zenodo.8125613}{https://doi.org/10.5281/zenodo.8125613}.

\appendix
\section{III-V Ternary Alloy Parameter Calculations}\label{app:iii_v_params}  %
Following the standard linear and quadratic interpolation schemes for III-V
ternary alloys \AxBxC{x}{1-x}, with composition parameter $x$ and in terms of experimentally measured values of their binary constituents, AB and BC, we have the lattice constant $a$, energy gap $E$, and effective mass at the $\Gamma$ point $m^{\Gamma\ast}$ as~\cite{Adachi2017}                    
    \begin{eqnarray}
        a_{\text{\AxBxC{x}{1-x}}}  &= x a\textsubscript{AC} + (1 - x) %
        a_{\text{BC}}, \label{eq:a_AxB1xC} \\                                   
        E_{\text{\AxBxC{x}{1-x}}} &= x E_{\text{AC}} + (1 - x) %
        E_{\text{BC}} + x (1 - x) E_{\text{AB}}, %
        \label{eq:E_AxB1xC} \\                                                  
        m^{\Gamma\ast}_{\AxBxC{x}{1-x}} &= x                                           
        m^{\Gamma}_{\text{AC}} %
        + (1 - x) m^{\Gamma}_{\text{BC}} + x (1 - x) %
        m^{\Gamma}_{\text{AB}}. \label{eq:mgamma_AxB1xC}                        
    \end{eqnarray}                                                              
Similarly, the hole effective masses follow from a quadratic interpolation      
scheme of the AB, AC binary components as computed from a spherical band        
approximation of the valence band edge~\cite{Adachi2017}
    \begin{eqnarray}
      m_{\mathrm{p,dos}} = \left(m_{\mathrm{lh}}^{3/2} + m_{\mathrm{hh}}^{3/2}\right)^{2/3}, %
      \label{eq:mp_dos} \\                                                      
      m_{\mathrm{p, c}} = \frac{m_{\mathrm{lh}}^{5/2} + m_{\mathrm{hh}}^{5/2}}{m_{\mathrm{p, dos}}}, %
      \label{eq:mp_c} \\                                                        
      m_{\mathrm{p, c, \AxBxC{x}{1-x}}} = %
      x m_{\mathrm{p, c, AC}} %
      + (1-x) m_{\mathrm{p, c, BC}}, \label{eq:mpc_AxB1xC} \\                     
      m_{\mathrm{p, dos, \AxBxC{x}{1-x}}} = %
      x m_{\mathrm{p, dos, AC}} %
      + (1-x) m_{\mathrm{p, dos, BC}}. \label{eq:mpdos_AxB1xC}             
    \end{eqnarray}
We recognize that the spherical band approximation may not apply to the III-V materials in our study, but it gives an estimate for density of states and conduction band effective masses that are inputs to the COMSOL Semiconductor Module materials models.

To estimate the conduction band offsets between the \InxAlxAs{x}{1-x} and \InxGaxAs{x}{1-x} layers, we followed another interpolation scheme that computes the absolute
conduction band edges $E_{\mathrm{c}}$ using experimentally measured parameters of InAs, AlAs, and GaAs\cite{Krijn1991}                              
   \begin{eqnarray}
       E_{\mathrm{c}} &= E\textsubscript{v,avg} + \frac{\Delta_0}{3} +  E_{\mathrm{g}} + \dEchy,%
       \label{eq:Ec} \\                                                         
       \Delta E_{\mathrm{c}} &= E_{\mathrm{c}}^B - E_{\mathrm{c}}^A, \label{eq:dEc}
   \end{eqnarray}                                                               
where $E_{\mathrm{v,avg}}$ is the average valence band edge, $\Delta_0$ is the    
spin-orbit splitting in the absence of strain, $E_{\mathrm{g}}$ is the band gap energy, and
$\Delta E_{\mathrm{c}}^{\mathrm{hy}}$ is the shift of the conduction band edge due to
hydrostatic strain. These parameters are calculated from the following          
expressions with coefficients $C_{ij}$ read-off from Table 3 compiled by Krijn~\cite{Krijn1991}
    \begin{eqnarray}                                                            
        E\textsubscript{v, avg} &= \sum_{i=1}^2 %
        C_{i0} (E\textsubscript{v,avg})x^i, \label{eq:Evavg} \\                
        \Delta_0 &= \sum_{i=1}^2 C_{i0} (\Delta_0) x^i,\label{eq:delta0} \\    
        \dEchy &= %
        \frac{\Delta a(x)}{a(x)} \sum_{i=0}^1 %
        C_{i0}(\dEchy) x^i, \label{eq:dEchy}\\                                  
        \Delta a(x) &= a_0 - a(x). \label{eq:delta_a}                          
    \end{eqnarray}
%
\section{Charge--Charge Interaction Matrix Element Derivation}\label{app:charge_matrix}
    \begin{figure}[!ht]
        \centering
        \includegraphics[width=0.6\textwidth]{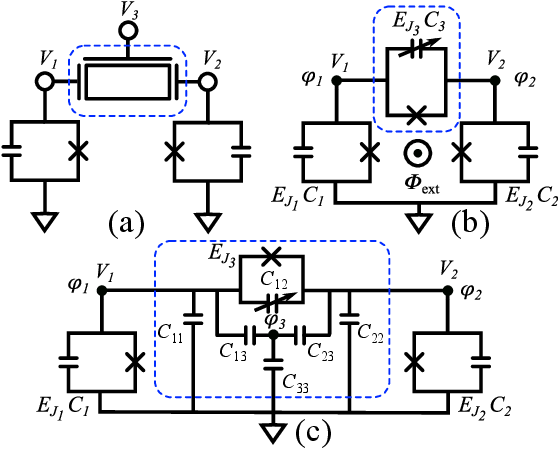}
        \caption{Coupler circuit models. (a) The 2DEG coupler compact representation with a single gate and a pair of transmon qubits compared with (b) the simplified circuit used in the derivation of the charge-charge interaction matrix in the main text. (c) Parasitic capacitance circuit model of the coupler and two transmons; capacitances taken from (\ref{eq:C_c_acdc}) and (\ref{eq:C_d_acdc}).}
        \label{fig:circuit_phases}
    \end{figure}
Starting from the two transmon circuit coupled by a voltage-controlled Josephson junction (our 2DEG coupler) in figure~\ref{fig:circuit_phases} (b), with phases $\varphi_1, \varphi_2, \varphi_3$ referring to the left, right, and coupling junctions, respectively, we have~\cite{Orlando1999}
    \begin{eqnarray}
        \varphi_1 - \varphi_2 + \varphi_3 = -\frac{2\pi}{\Phi_0} \Phiext, \label{eq:phase_loop_rule}
    \end{eqnarray}
where $\Phiext$ is the flux threading the loop formed by the three Josephson junctions as in a typical flux qubit circuit. For a finite $\Phiext$, the potential energy $U$ is given by
    \begin{eqnarray}
        U(\mathbf{\varphi}) &= \sum_j E_{\mathrm{J}_j}(1 - \cos\varphi_j) \nonumber \\
        &= E_{\mathrm{J}_1}(1 - \cos\varphi_1) + E_{\mathrm{J}_2}(1 - \cos\varphi_2) \nonumber \\
        &+ E_{\mathrm{J}_3}(1 - \cos(\varphi_2 - \varphi_1 - 2\pi\Phiext/\Phi_0)), \label{eq:U_phi}
    \end{eqnarray}
with the signs on the phases following figure~\ref{fig:circuit_phases} (b), preserving the conventions chosen in.~\cite{Orlando1999} We order the phases in a single column vector as
    \begin{eqnarray}
        \mathbf{\varphi} = 
        \begin{pmatrix}
        \varphi_1 \\
        \varphi_2
        \end{pmatrix}. \label{eq:def_phi_vector}
    \end{eqnarray}
Setting $\Phiext$ = 0, we compute the kinetic energy $T$ by using the Josephson equation relating the voltages at nodes with $k=\left\{1,\ 2\right\}$, $V_k = (\Phi_0/2\pi)\dot{\varphi}_k$ and the definition of $T$ in terms of $\dot{\varphi}_k$
    \begin{eqnarray}
        T &= \frac{1}{2}\left(C_1 V_1^2 + C_2 V_2^2 + C_3 V_3^2\right) \nonumber \\
          &= \frac{1}{2}\left(\frac{\Phi_0}{2\pi}\right)^2\left(C_1\dot{\varphi}^2_1 + C_2\dot{\varphi}^2_2 
          + C_3\left(\dot{\varphi}_2 - \dot{\varphi}_1\right)^2\right) \nonumber \\
          &= \frac{1}{2}\left(\frac{\Phi_0}{2\pi}\right)^2 \dot{\mathbf{\varphi}}^T \mathbf{C} \dot{\mathbf{\varphi}}, \label{eq:T_phi_dot}
    \end{eqnarray}
and reading off the capacitance matrix
    \begin{eqnarray}
        \mathbf{C} =
        \begin{pmatrix}
        C_1 + C_3 & -C_3 \\
        -C_3 & C_2 + C_3
        \end{pmatrix}. \label{eq:def_c_matrix_qq}
    \end{eqnarray}
Relating the total capacitances (both the intrinsic junction and external capacitance, commonly referred to as $C_{\Sigma}$~\cite{Schuster2007}) shunting the junctions, $C_1,\ C_2$, to the anharmonicities extracted from the EPR calculations, we have, from the asymptotic expressions derived by Koch~\etal~\cite{Koch2007}
    \begin{eqnarray}
        C_k &= \frac{e^2}{2E_{\mathrm{C}}} \simeq -\frac{e^2}{2\alpha_k} \label{eq:alpha_to_CJ}
    \end{eqnarray}
and we take $C_3$ = $C_{12}(V_{\mathrm{g}})$, the gate voltage-dependent capacitance across the 2DEG coupler. The classical Lagrangian $\mathcal{L}$ and Hamiltonian $\mathcal{H}$ associated with the kinetic and potential energies above, then read~\cite{Orlando1999}
    \begin{eqnarray}
        \mathcal{L(\pmb{\varphi}, \pmb{\dot{\varphi}})} &= T - U \nonumber \\
        &= \frac{1}{2} \left(\frac{\Phi_0}{2\pi}\right)^2 \dot{\pmb{\varphi}}^T \mathbf{C} \dot{\pmb{\varphi}} 
        - \sum_j E_{\mathrm{J}_j}(1 - \cos\varphi_j) \label{eq:def_classical_lagrangian} \\
        \mathcal{H} &= \mathbf{P}^T \dot{\pmb{\varphi}} - \mathcal{L} \nonumber \\
        &= \frac{1}{2}\mathbf{Q}^T \mathbf{C}^{-1} \mathbf{Q} + U(\pmb{\varphi}) \label{eq:def_classical_hamiltonian} \\
        P_j &= \frac{\partial \mathcal{L}}{\partial \dot{\varphi_j}} 
        = \left(\frac{\Phi_0}{2\pi}\right)^2 \sum_kC_{jk}\dot{\varphi}_k, \ \mathbf{Q} = \frac{2\pi}{\Phi_0} \mathbf{P} \label{eq:momenta}
    \end{eqnarray}
We take the form of the quantized Hamiltonian to be the same as the classical one in (\ref{eq:def_classical_hamiltonian}) with classical variables promoted to operators, and identify the charge-charge matrix elements as $e^2[\mathbf{C}^{-1}]_{ij}/2$. Similarly, we write the Lagrangian and identify the capacitance matrix corresponding to the parasitic capacitance model given by the circuit in Fig.~\ref{fig:circuit_phases} (c) as
    \begin{eqnarray}
        \mathcal{L} &= \frac{1}{2}
        \left( \frac{\Phi_0}{2\pi} \right)^2
        \left[
          (C_1 + C_{11}) \dot{\varphi}^2_1
        + (C_2 + C_{22}) \dot{\varphi}^2_2 \right. \nonumber \\
        &+ \left. C_{33} \dot{\varphi}^2_3
        + C_{12}(\dot{\varphi}_2 - \dot{\varphi}_1)^2\right. \nonumber \\
        &+ \left. C_{13}(\dot{\varphi}_3 - \dot{\varphi}_1)^2
         + C_{23}(\dot{\varphi}_2 - \dot{\varphi}_3)^2
        \right] - U(\varphi) \label{eq:lagrangian_parasitic}
    \end{eqnarray}
        \begin{eqnarray}
            \mathbf{C} &=
            \begin{pmatrix}
            \widetilde{C}_{11} & -C_{12} & -C_{13} \\
            -C_{12} & \widetilde{C}_{22} & -C_{23} \\
            -C_{13} & -C_{23} & \widetilde{C}_{33}
            \end{pmatrix}, \label{eq:cmatrix_parasitic}
        \end{eqnarray}
%
where $\widetilde{C}_{11}=C_1 + C_{11} + C_{12} + C_{13}$, $\widetilde{C}_{22}=C_{2} + C_{22} + C_{12} + C_{23}$, and $\widetilde{C}_{33}=C_{13} + C_{23} + C_{33}$.

\bibliography{references.bib}

\end{document}